\documentclass[twoside,11pt]{article}
\pdfoutput=1
\usepackage{jmlr2e_arxiv}

\usepackage{url}
\usepackage{natbib}
\usepackage{amsmath}
\usepackage{amssymb}
\usepackage{bm}
\usepackage{subfigure}
\usepackage{graphicx}
\usepackage{enumerate}

\usepackage[hyperindex,breaklinks]{hyperref}

\newcommand{\f}{\bm{\mathrm{f}}}
\newcommand{\x}{\bm{\mathrm{x}}}
\newcommand{\y}{\bm{\mathrm{y}}}
\newcommand{\z}{\bm{\mathrm{z}}}

\newcommand{\h}{\bm{\mathrm{h}}}

\newcommand{\uu}{\bm{\mathrm{u}}}
\newcommand{\vv}{\bm{\mathrm{v}}}
\newcommand{\bb}{\bm{\mathrm{b}}}
\DeclareMathOperator{\Gam}{Gam}

\ShortHeadings{Laplace Approximation for Logistic GP Density
  Estimation}{Riihim\"{a}ki and Vehtari}
\firstpageno{1}

\begin{document}

\thispagestyle{empty}

\title{Laplace Approximation for Logistic Gaussian
  Process\\Density Estimation and Regression}

\author{\name Jaakko Riihim\"{a}ki \email jaakko.riihimaki@aalto.fi \\
       \name Aki Vehtari \email aki.vehtari@aalto.fi \\
       \addr Department of Biomedical Engineering and Computational Science\\
       Aalto University\\
       P.O. Box 12200\\
       FI-00076 Aalto\\
       Finland}


\maketitle

\begin{abstract}

  Logistic Gaussian process (LGP) priors provide a flexible
  alternative for modelling unknown densities. The smoothness
  properties of the density estimates can be controlled through the prior
  covariance structure of the LGP, but the challenge is the
  analytically intractable inference. In this paper, we present
  approximate Bayesian inference for LGP density estimation in a grid
  using Laplace's method to integrate over the non-Gaussian posterior
  distribution of latent function values and to determine the
  covariance function parameters with type-II maximum a posteriori
  (MAP) estimation. We demonstrate that Laplace's method with MAP is
  sufficiently fast for practical interactive visualisation of 1D and
  2D densities. Our experiments with simulated and real 1D data sets
  show that the estimation accuracy is close to a Markov chain Monte
  Carlo approximation and state-of-the-art hierarchical infinite
  Gaussian mixture models. We also construct a reduced-rank
  approximation to speed up the computations for dense 2D grids, and
  demonstrate density regression with the proposed Laplace approach.

\end{abstract}


\section{Introduction}
\label{sec:introduction}



Logistic Gaussian process (LGP) priors provide a flexible alternative
for modelling unknown densities \citep{Leonard:1978}.
With the LGP models densities can be estimated without restricting to
any specific parameterized form and the smoothness properties of
estimates can be controlled through the prior covariance structure.
%
%
The challenge with the LGP model is the analytically intractable
inference that results from the normalization term required to
construct the prior distribution over density functions.
%
This paper focuses on finite dimensional approximations, where the
Gaussian process and the integral required to ensure normalization are
evaluated in a grid as described by \citet{Tokdar:2007}.
%
%
%
The theoretical properties of LGP are discussed by
\citet{Tokdar+Ghosh:2007} who establish conditions for posterior
consistency of LGP density estimation and by
%
\citet{vandervaart2009} who consider posterior convergence rates.
%
%
\citet{Tokdar:2007} shows that as 
the spacing between the grid points decreases,
%
the Kullback--Leibler divergence from an infinite-dimensional model to
the finite-dimensional approximation converges to zero. Related
considerations about a finite element approach for LGP density
estimation are also presented by \citet{Griebel+Hegland:2010}.
%
%
%
Densities are often estimated on bounded intervals, although the
estimation can be extended to
unbounded intervals by transforming them into bounded intervals as
proposed by \citet{Tokdar:2007}.

\citet{Tokdar:2007} integrated over the latent values with
Metropolis--Hastings sampling and over the potential grid point sets
with Metropolis--Hastings--Green sampling.
The purpose of the latter part is to keep the number of active grid
points small and automatically find the most important grid point
locations.
%
%
By replacing the sampling with an analytic approximation, inference
can be made faster even if a fixed and finer grid is assumed.
%
%
In this paper, we consider Laplace's method (LA) to approximate the
posterior inference for LGP density estimation in a grid.
%
Our objective is to obtain an accurate and quick approximation that
enables practical estimation of densities by focusing on efficient
ways to approximate Bayesian inference for LGP density estimation.




The proposed LA approach is related to other approaches in the
literature. Given fixed covariance function parameters, the LA
approximation for the posterior distribution involves 
finding the posterior mode, which is equivalent to a penalized maximum
likelihood estimator considered by \citet{Leonard:1978} and
\citet{Thorburn:1986}.
The marginal likelihood can be approximated with LA, which
enables 
a fast gradient-based type-II maximum a posteriori (MAP) estimation of
the covariance function parameters, or alternatively,
the posterior distribution can be integrated over.
%
%
%
\citet{Lenk:1991} uses a truncated Karhunen--Loeve representation and
derives the moments of the density process, from which it is also
possible to obtain a marginal likelihood approximation. Lenk evaluates
the marginal likelihood of hyperparameter values in a grid, while we
use gradients and BFGS quasi-Newton optimization or Markov chain
Monte Carlo (MCMC) integration. 
In Lenk's approach the mean of the density process is guaranteed to be
positive, but there is no such guarantee for the whole posterior.
Also the spectral representation restricts the selection of covariance
functions. 
\citet{Lenk:2003} refines his approach with a Fourier series expansion
and MCMC sampling of the hyperparameters.

The computational complexity of the proposed LA approach is dominated
by the covariance matrix operations, which scale in a straightforward
implementation as $\mathcal{O}(m^3)$, where $m$ is the number of grid
points. However, because of the discretization, the computational
complexity is independent of the number of observations.
%
Applying the proposed approach with a default grid size 400, 1D and
2D density estimation takes one to two seconds, which facilitates
interactive
visualization of data with density estimates and violin plots (see,
e.g., Figures \ref{fig:sim_examples}--\ref{fig:real_comp} and
\ref{fig:2d_examples}).
%
Additionally, we consider fast Fourier transform (FFT) to speed up
the computations with an even grid and stationary covariance
functions. To avoid the cubic computational scaling in $m$ with dense
2D grids, we also exploit Kronecker product computations to obtain a
reduced-rank approximation of the exact prior covariance structure.
%
%
%

The number of grid points grows exponentially with the data dimension
$d$.
%
Although the Kronecker product approach is suitable for reducing the
computation time when $d>1$, the exponentially increasing number of
latent function values makes the proposed approach impractical for
more than three or four dimensions.
%
%
\citet{Adams+etal:2009a} propose an alternative GP approach called
Gaussian process density sampler (GPDS) in which the numerical
approximation of the normalizing term in the likelihood is avoided by
a conditioning set and an elaborate rejection sampling method.
%
%
The conditioning set is generated by the algorithm, which
automatically places $m_s$ points where they are needed, making the
estimation in higher-dimensional spaces easier. However, the
computational complexity of GPDS 
scales as $\mathcal{O}((n+m_s)^3)$, where $n$ is the number of data points.

The main contribution of this paper is the construction of the quick
approximation for the LGP density estimation and regression by
combining various ideas from the literature. Using Laplace's method to
integrate over the latent values, we avoid the slow mixing of MCMC. We
present FFT and reduced-rank based speed-ups tailored for LGP with
LA. We demonstrate that LA can be further improved by rejection and
importance sampling. We also show that using the type-II MAP
estimation for two to three hyperparameters gives good results compared to
the integration over the hyperparameters.
%
%
%
%

In the next section, we review the basics of the logistic Gaussian
processes. 
In Section \ref{sec:appr-infer}, we present the LA approach for LGP
density estimation. We also introduce the additional approximations
for speeding up the inference and consider briefly a similar LA
approach for LGP density regression.
In Section \ref{sec:experiments}, we demonstrate the LA approach with
several experiments, and compare it against MCMC and hierarchical
infinite Gaussian mixture models by \citet{Griffin:2010}.


\section{Density Estimation with Logistic Gaussian Process Priors}
\label{sec:dens-estim-with}

%
We consider the problem of computing a density estimate $p(\x)$ given
$n$ independently drawn $d$-dimensional data points $\x_1,\ldots,\x_n$
from an unknown distribution in a finite region $\mathcal{V}$ of
$\mathbb{R}^d$.
In this paper, we focus only on $d\in\{1,2\}$.
To find an estimate $p$ for the unknown density, we can maximize the
following log-likelihood functional
\begin{equation}
L(p)=\sum_{i=1}^n \log p(\x_i)
\end{equation}
with the constraints
$\int_{\mathcal{V}}p(\x)d\x=1$,
and $p(\x)\ge0$ for $\x\in \mathcal{V}$. The limiting solution leads
to a mixture of delta functions located at the observations
\citep{Leonard:1978}, which is why we need to set prior beliefs about
the unknown density to obtain more realistic estimates. 
%

To introduce the 
constraints of the density being non-negative and that its integral
over 
$\mathcal{V}$ 
is equal to one, we employ the logistic density transform~\citep{Leonard:1978} 
\begin{equation}\label{eq_logdens}
p(\x)=\frac{\exp(f(\x))}{\int_{\mathcal{V}}\exp(f(\bm{s}))d\bm{s}},
\end{equation}
where $f$ is an unconstrained latent function.
To smooth the density estimates,
we place a Gaussian process prior for $f$, 
which enables us to set the prior assumptions about the smoothness
properties of the unknown density $p$ via the covariance structure of
the GP prior.

We assume a zero-mean Gaussian process
$g(\x)\sim \mathcal{GP}\left(0,\kappa(\x,\x')\right)$,
where the covariance function is denoted with $\kappa(\x,\x')$ for a pair
of inputs $\x$ and $\x'$. An example of a widely used covariance
function is the stationary squared exponential covariance function
\begin{eqnarray}\label{cf_sexp}
\kappa(\x,\x')=\sigma^2\exp\left(-\frac{1}{2}\sum_{k=1}^{d}l_k^{-2}(x_k-x'_k)^2\right),
\end{eqnarray}
where the hyperparameters $\theta=\{\sigma^2,l_1,\ldots,l_d \}$ govern the
smoothness properties of $f$ \citep{rasmussen2006}. 
The length-scale hyperparameters $l_1,\ldots,l_d$ control how fast the
correlation decreases in different dimensions, and 
$\sigma^2$ is a magnitude hyperparameter.

\begin{figure*}[t]
  \centering
  \includegraphics[scale=0.85,clip]{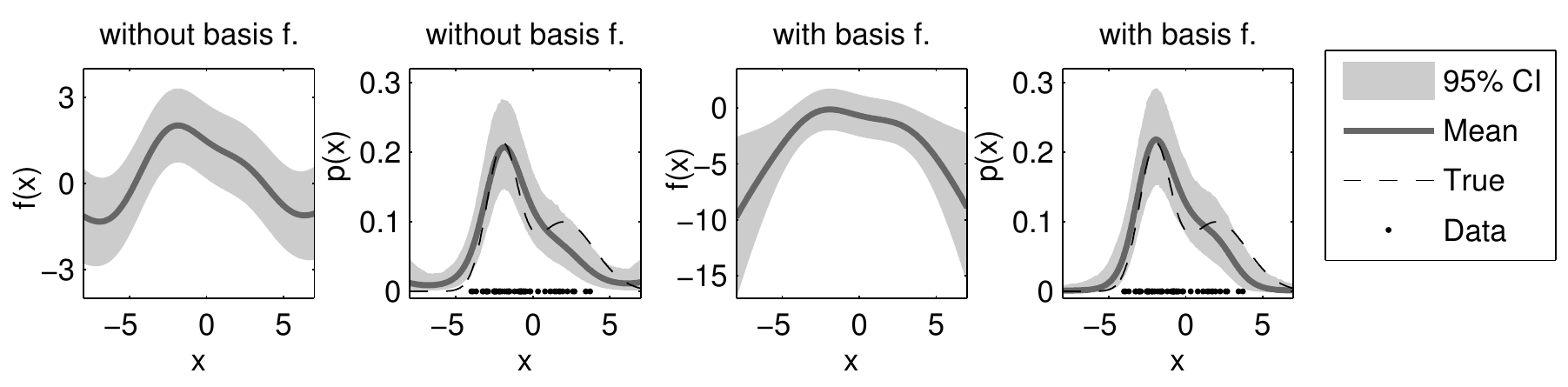}
  \caption{An illustration of the logistic Gaussian process density
    estimation with and without the basis functions. The first plot
    visualizes the posterior latent function without the basis
    functions, and the second plot shows the corresponding density
    estimate with the logistic density transform \eqref{eq_logdens}
    given 50 observations from the mixture of two Gaussians:
    $\frac{1}{2}\mathcal{N}(-2,1)+\frac{1}{2}\mathcal{N}(2,2^2)$.  The
    third plot visualizes the posterior latent function with the
    second-order polynomials as the basis functions, and the fourth
    plot shows the corresponding density estimate. The hyperparameter
    values of the squared exponential covariance function
    \eqref{cf_sexp} are in this example $\sigma^2=1$ and
    $l_1=\frac{1}{2}$.}
\label{fig:pictorial}
\end{figure*}

For the latent function $f$ in Equation~\eqref{eq_logdens}, we assume
the model $f(\x) = g(\x)+\h(\x)^T \bm{\beta}$, where the GP prior is
combined with the explicit basis functions $\h(\x)$.  Regression
coefficients are denoted with $\bm{\beta}$, and by placing a Gaussian
prior $\bm{\beta}\sim \mathcal{N}(\bb,B)$ with the mean $\bb$ and
the covariance $B$, the parameters $\bm{\beta}$ can be integrated out from
the model, which results in the following GP prior for $f$ \citep{ohagan1978,rasmussen2006}:
\begin{equation}\label{gp_process}
  f(\x)\sim \mathcal{GP}\left(\h(\x)^T\bb,\kappa(\x,\x')+\h(\x)^TB\h(\x')\right).
\end{equation}
We assume the second-order polynomials for the explicit basis
functions.
%
We use $\h(\x)=[x_1, \, x_1^2]^T$ in 1D, and $\h(\x)=[x_1, \, x_1^2,
\, x_2, \, x_2^2, \, x_1x_2]^T$ in 2D,
which leads to a GP prior that can favour density estimates where the
tails of the distribution go eventually to zero. The effect of the
basis functions is demonstrated in an illustrative 1D example of
Figure \ref{fig:pictorial}.

We discretize 
$\mathcal{V}$ into
$m$ subregions (or intervals in 1D),
and collect the coordinates of the subregions into an $m \times d$
matrix $X$, where the $i$'th row denotes the center point of the
$i$'th subregion.
Given $X$, the GP prior (\ref{gp_process}) results in the Gaussian
distribution over the latent function values
\begin{equation}\label{gp_prior}
  p(\f|X,\theta)=\mathcal{N}(\f|H\bb,K+HBH^T),
\end{equation}
where 
$\f$ is a column vector of $m$ latent values associated with
each subregion. 
The entries of the $m \times m$ covariance matrix $K$
%
%
are determined by the input points $X$ and the covariance function.
%
%
The matrix $H$, of size $m\times 2$ in 1D and $m\times 5$ in 2D,
contains the values of the fixed basis functions evaluated at $X$.
We assume a weakly informative prior distribution for the regression
coefficients by fixing the mean $\bb=\bm{0}$ (a zero vector of a
length $2$ in 1D and $5$ in 2D), and the covariance $B=10^2 I$, where
$I$ is the identity matrix of a size $2\times 2$ (in 1D) and $5\times
5$ (in 2D). 
The regression coefficient for the quadratic term $x_k^2$ should be
negative in order to make the tails of a distribution to go towards
zero. However, the prior distribution does not force the negativity of
the coefficient. In our experiments, the posterior of the coefficient
was clearly below zero, but in our implementation it is
also 
possible to force the negativity of the coefficient by rejection
sampling (in 1D).

After the discretization, the log-likelihood contribution of an
observation belonging to the $i$'th subregion can be written as
\begin{equation}\label{eq_lik}
  L_i=\log\left(\frac{w_i\exp(f_i)}{\sum_{j=1}^m w_j\exp(f_j)}\right),
\end{equation}
where the latent value $f_i$ is associated with the $i$'th subregion.
Throughout this paper, we assume a regular grid,
and therefore the weights 
$w_1,\ldots,w_m$ have all the same value and can be omitted from
\eqref{eq_lik}. The number of observations that fall within the $i$'th
subregion is denoted with $y_i$ and all the count observations with an
$m\times 1$ vector $\y$.
%
%
The overall log-likelihood contribution of the $n$ observations is
given by 
\begin{eqnarray}\label{likelih_logitgp}
\log p(\y|\f)
= \y^T\f - n\log \left(\sum_{j=1}^m \exp(f_j)\right).
\end{eqnarray}


The prior \eqref{gp_prior} and the likelihood \eqref{likelih_logitgp}
are combined by 
Bayes' rule, which results in the conditional posterior distribution
of the latent values 
\begin{equation}
p(\f|X,\y,\theta)=\frac{1}{Z}p(\f|X,\theta)p(\y|\f),
\end{equation}
where 
$Z=p(\y|X,\theta)=\int p(\f|X,\theta)p(\y|\f) d\f$ is the marginal
likelihood. Due to the non-Gaussian likelihood
\eqref{likelih_logitgp}, the posterior distribution is also
non-Gaussian,
and approximate methods are needed to integrate over $\f$.


\section{Approximate Inference}
\label{sec:appr-infer}


In this section, we discuss the implementation issues of Laplace's
method for logistic GP density estimation.  In Section
\ref{sec:laplace-approximation}, we present an efficient approach for
the mode finding and computation of the marginal likelihood and
predictions. Further speed-ups are obtained by using the fast Fourier
transform for matrix-vector multiplications and by Kronecker
product and reduced-rank approximations suitable for $d>1$ cases with
dense grids (large $m$).
In Section \ref{sec:density_regression}, we consider the LA approach
for logistic GP density regression, and in Section~\ref{sec:mcmc}, we
give a brief description of inference with MCMC.

%




\subsection{Inference with the Laplace Approximation}
\label{sec:laplace-approximation}

%

Our approach resembles the Laplace approximation for GP classification
\citep{williams1998,rasmussen2006} and GP point process intensity
estimation \citep{Cunningham+Shenoy+Sahani:2008}, but the
implementation is different because in LGP each term in the likelihood
\eqref{likelih_logitgp} depends on all the latent values $\f$.

The Laplace approximation is based on a second-order Taylor expansion for
$\log p(\f|X,\y,\theta)$ around the posterior mode $\hat{\f}$, which
results in the Gaussian approximation
\begin{equation}
  q(\f|X,\y,\theta)=\mathcal{N}(\f|\hat{\f},\Sigma),
\end{equation}
where $\hat{\f}=\mathrm{arg} \max_{\f}p(\f|X,\y,\theta)$. The
covariance matrix is given by
\begin{equation}\label{post_cov}
\Sigma=(C^{-1}+W)^{-1},
\end{equation}
where $C=K+HBH^T$ and $W=-\nabla \nabla_{\f} \log
p(\y|\f)|_{\f=\hat{\f}}$. 
The likelihood (\ref{likelih_logitgp}) leads to a full matrix $W$ with
the following structure\footnote{We use the following notation:
  diag($\bm{\mathrm{w}}$) with a vector argument means a diagonal
  matrix with the elements of the vector $\bm{\mathrm{w}}$ on its
  diagonal, and diag($W$) with a matrix argument means a column vector
  consisting of the diagonal elements of the matrix $W$.}
\begin{equation}\label{lik_structure}
  W=n(\mathrm{diag}(\uu)-\uu\uu^T),
\end{equation}
where the non-negative entries of the vector $\uu$
are given by
\begin{eqnarray}\label{lik_u}
u_i=\frac{\exp(f_i)}{\sum_{j=1}^m \exp(f_j)}.
\end{eqnarray}
%
%
%
In the implementation, forming the full matrix $W$ can be avoided by
using the vector $\uu$ and the structure \eqref{lik_structure}.
Similarly to multiclass classification with the softmax likelihood
\citep{williams1998,rasmussen2006}, $W$ is positive semidefinite, and
because $C$ is positive definite, $p(\f|X,\y,\theta)$ has a unique
maximum. 




\subsubsection{Newton's Method for Finding the Mode}\label{sec:newton}

We use Newton's method for finding the mode $\hat{\f}$. At each
iteration, we need to compute
\begin{equation}\label{newton_update}
\f^{\mathrm{new}}=(C^{-1}+W)^{-1}\vv,
\end{equation}
where $\vv=W\f+\nabla_{\f}\log p(\y|\f)$. 
%
%
To increase the numerical stability of the computations, we use the
factorization $W=RR^T$, where
\begin{eqnarray}\label{lik_structure_R}
  R=\sqrt{n}((\mathrm{diag}(\uu))^{\frac{1}{2}}-\uu\uu^T(\mathrm{diag}(\uu))^{-\frac{1}{2}}).
\end{eqnarray}
Instead of a direct implementation of \eqref{newton_update}, we can
apply the matrix inversion lemma \citep[see, e.g.,][]{Harville:1997}
to write the Newton step in a numerically more preferable way as
\begin{eqnarray}\label{newton_robust}
\f^{\mathrm{new}}=C(I_m-R(I_m+R^TCR)^{-1}R^TC)\vv.
\end{eqnarray}
%
%
%
The inversion of Equation~\eqref{newton_robust} can be computed by
solving $\z$ from the linear system $(I_m+R^TCR)\z=R^TC\vv$ with the
conjugate gradient method.
%
%
%
%
%
As discussed, for example, by \citet{Cunningham+Shenoy+Sahani:2008},
a stationary covariance function and evenly spaced grid points lead to
a Toeplitz covariance matrix which can be embedded in a larger
circulant matrix enabling efficient computations by using the fast
Fourier transform.
%
%
With a Toeplitz covariance matrix $K$, we can achieve a small speed-up
with larger grid sizes in the evaluation of the Newton step
\eqref{newton_robust} because all the multiplications of $K$ and any
vector can be done efficiently in the frequency domain.
By using FFT, these matrix-vector multiplications become convolution
operations for which we are required only to form a single row of the
embedded circulant matrix instead of the full matrix $K$.
The rest of the matrix-vector multiplications in \eqref{newton_robust}
are fast because the matrix $H$ has only two (in 1D) or five (in 2D)
columns, and instead of forming the full matrix $R$, we can use the
vector $\uu$ and exploit the structure of $R$ from Equation
\eqref{lik_structure_R}.

%










\subsubsection{The Approximate Marginal Likelihood and Predictions}


Approximating the integration over $\f$ with the LA approximation
enables fast gradient-based type-II MAP estimation for choosing the
values for the covariance function hyperparameters.
After finding $\hat{\f}$, the approximate log marginal likelihood can
be evaluated as
\begin{eqnarray}\label{approximate_ml}
  \log p(\y|X,\theta)\approx \log q(\y|X,\theta)=-\frac{1}{2}
  \hat{\f}^TC^{-1}\hat{\f}+\log p(\y|\hat{\f})-\frac{1}{2}\log|I_m+R^TCR|
\end{eqnarray}
\citep[see, e.g.,][]{rasmussen2006}. The first and second terms of
\eqref{approximate_ml} are fast to compute by using the results from
Newton's algorithm, but the determinant term is more difficult to
evaluate.

\citet{Cunningham+Shenoy+Sahani:2008} show that for the point
process intensity estimation with the GP priors, the evaluation of a
corresponding determinant term can be done efficiently by exploiting a
low-rank structure of the observation model.
In density estimation $W$ has rank $m-1$ due to the normalization
over 
$\mathcal{V}$,
and a similar low-rank representation as in intensity estimation
cannot be used for $W$.
Therefore, we form the full $m\times m$ matrix and compute the
Cholesky decomposition that scales as $\mathcal{O}(m^3)$.
Although this is computationally burdensome for large $m$, it is
required only once per each evaluation of the marginal likelihood
after the convergence of Newton's algorithm.  Typically, in 1D cases
with the number of intervals $m$ being less than one thousand, the LA
approach is sufficiently fast for practical purposes (in our
implementation the default size for $m$ is 400).  However, the
computations can become restrictive in 2D with large $m$.  In such
cases, we consider a reduced-rank approximation for the prior
covariance $K$ to find a faster way to compute the determinant term
and the approximate posterior covariance, as will be discussed in
Section \ref{sec:low-rank}.
%
%
The gradients of the log marginal likelihood with respect to the
hyperparameters $\theta$ can be computed by evaluating explicit and
implicit derivatives similarly as shown by \citet{rasmussen2006}.

We place a prior distribution $p(\theta)$ for the hyperparameters to
improve the identifiability of the ratio of the magnitude and the
length-scale parameters of the covariance function.
%
%
We assume a weakly informative half Student-$t$ distribution, as
recommended for hierarchical models by \citet{Gelman:2006}, with 
one degree of freedom and a variance that is equal to ten in 1D and
thousand in 2D for
$\sigma$ (magnitude). For $l_1,\ldots,l_d$ (length-scales), we assume
otherwise the same prior but with a variance that is equal to one.
The MAP estimate for the hyperparameters is 
found by maximizing the approximate marginal posterior
$p(\theta|\y,X)\propto q(\y|X,\theta)p(\theta)$, in which we use the
BFGS quasi-Newton optimization.
In addition to the MAP estimate, we also approximate the integration
over the hyperparameters with the central composite design (LA-CCD-LGP)
scheme, similarly as proposed by \citet{rue2009}.
%

To compute the joint posterior predictive distribution, we marginalize
over the latent values by Monte Carlo using 8000 latent samples drawn
from the multivariate normal posterior predictive distribution. This
sampling is needed only once after the MAP estimate for the
hyperparameters has been found,
%
%
and time consumed in this step is negligible compared to the
other parts of the computations.
In many cases, we have observed that the posterior weights of the
quadratic terms $x_k^2$ of the basis function are automatically small
and thus the effect of basis functions is not strong for densities not
well presented by them.
We consider an optional rejection sampling step based on finite
differences to enforce the tails to go to zero (in 1D), if necessary
(not used if density is defined to be bounded).
%

To improve the Gaussian approximation of the posterior distribution,
we also consider an additional importance sampling step
(LA-IS-LGP). Following \citet{Geweke:1989} we use the
multivariate split Gaussian density as an approximation for the exact
posterior. The multivariate split Gaussian is based on the posterior
mode and covariance, but the density is scaled along the principal
component axes (in positive and negative direction separately)
adaptively to match to the skewness of the true distribution
\citep[see also][]{Villani+Larsson:2006}. To further improve the
performance the discontinuous split Gaussian used by
\citet{Geweke:1989} was replaced with a continuous version.
We observed that scaling is not needed along the principal component
axis corresponding to the smallest eigenvalues. To speed up the
computation, we scale only along the first 50 principal component
axis.
The importance sampling step took approximately additional
0.3 seconds, whereas direct sampling from the Gaussian posterior
distribution (to compute the predictions) took about 0.05 seconds.
Due to this small additional computational demand, we added the
importance sampling correction also in the experiments. In all the
experiments the estimated effective number of samples
\citep{Kong+Liu+Wong:1994} was reasonable, but we also included a
sanity check to the code and give a warning and use a soft
thresholding of the importance weights if the estimated effective
number of samples is less than $200$.
As a comparison, using scaled Metropolis--Hastings sampling to obtain
500 effectively independent samples from the latent posterior given
fixed hyperparameters took 24 minutes (similar inefficiency in mixing
was observed with other MCMC methods usually used for GPs).

\subsubsection{The Reduced-Rank Approximation for 2D Density
  Estimation}\label{sec:low-rank}

%
To speed up the inference with 2D grids when $m$ is large, we propose
a reduced-rank approximation that can be formed efficiently by
exploiting Kronecker product computations.
For separable covariance functions, the covariance structure of evenly
spaced grid points can be presented in a Kronecker product form $K=K_1
\otimes K_2$, where $K_k$ is an $m_k \times m_k$ matrix representing
the covariances between the latent values associated with $m_k$ grid
points in the $k$'th dimension.
%
For Kronecker products, many matrix operations scale efficiently: for
example, the determinant $|K_1 \otimes K_2|$ can be computed by using
the determinants of the smaller matrices $K_1$ and $K_2$ as
$|K_1|^{m_2}|K_2|^{m_1}$ \citep{Harville:1997}.
To compute 
the approximate marginal likelihood \eqref{approximate_ml}, we need to
evaluate the determinant term of a form $|I_m+R^T(K_1 \otimes
K_2+HBH^T)R|$. Unfortunately, the Kronecker product covariance
structure does not preserve due to the multiplication and summation
operations, leading to the unfavourable $\mathcal{O}(m^3)$
scaling.
%
%
%
%
%
However, we can exploit the Kronecker product $K_1 \otimes K_2$ to
obtain the eigendecomposition of $K$ efficiently, and then form a
reduced-rank approximation for $K$ by using only the largest
eigenvalues with the corresponding eigenvectors.
The idea of using the eigendecomposition to construct a reduced-rank
approximation for the covariance matrix has been previously mentioned
by \citet[][Chapter 8]{rasmussen2006}.
%
%
By denoting an eigenvalue of $K_1$ with $r_1$ and an eigenvector of
$K_1$ with $\vv_1$, and similarly, an eigenvalue of $K_2$ with $r_2$
and an eigenvector of $K_2$ with $\vv_2$, then, $r_1 r_2$ is an
eigenvalue of $K_1 \otimes K_2$ and $\vv_1 \otimes \vv_2$ an
eigenvector of $K_1 \otimes K_2$ corresponding to the eigenvalue $r_1
r_2$ \citep{Harville:1997}.
Thus, instead of computing the eigendecomposition of $K$, which is an
$\mathcal{O}(m^3)$ operation, we can compute the eigendecompositions
of $K_1$ and $K_2$, which scales as $\mathcal{O}(m_1^3+m_2^3)$, and
form a desired number of eigenvalues and eigenvectors with the
Kronecker product computations to obtain the reduced-rank
approximation for $K$.

We approximate the exact prior covariance with
\begin{eqnarray}\label{approx_prior_cov}
  K\approx VSV^T+\Lambda,
\end{eqnarray}
where $S$ is a diagonal matrix of size $s\times s$ having the $s$
largest eigenvalues on its diagonal and $V$ is an $m\times s$ matrix
consisting of the corresponding eigenvectors. Similarly to the fully
independent conditional (FIC) approximation
\citep{Snelson+Ghahramani:2006}, we use the exact full-rank diagonal
by setting $\Lambda=\mathrm{diag}(\mathrm{diag}(K)-
\mathrm{diag}(VSV^T))$ in \eqref{approx_prior_cov} to obtain a more
accurate approximation.  

With the approximate prior \eqref{approx_prior_cov}, the mode can be
found using Newton's method in a similar way as described in Section
\ref{sec:newton}.
%
To evaluate the marginal likelihood approximation, we need to compute
efficiently the determinant in \eqref{approximate_ml}. The determinant
term can be written as $n|A||\bm{1}^TD^{1/2}A^{-1}D^{1/2}\bm{1}|$,
where $\bm{1}$ is an $m\times 1$ column vector of ones,
$D=n(\mathrm{diag}(\uu))$ is a diagonal matrix, and
$A=I_m+D^{1/2}(\Lambda+\tilde{V}\tilde{S}\tilde{V}^T)D^{1/2}$. We have
defined $\tilde{V}=\begin{bmatrix}V & H \end{bmatrix}$, which is of
size $m \times (s+5)$, and $\tilde{S}=\begin{bmatrix} S & 0 \\ 0 & B
\end{bmatrix}$ of size $(s+5) \times (s+5)$.
To avoid forming any $m\times m$ matrix, $|A|$ can be evaluated by
applying the matrix determinant lemma \citep[see,
e.g.,][]{Harville:1997} and $A^{-1}$ by applying the matrix inversion
lemma.
With the prior \eqref{approx_prior_cov}, we can also compute the
approximate gradients of the marginal likelihood with respect to
hyperparameters without resorting to any $\mathcal{O}(m^3)$ matrix
operations.
%
%



For large $m$, we use the same fixed grid for the observations and
predictions. 
After we have found the MAP estimate for the hyperparameters, we need
to draw samples from $q(\f|X,\y,\theta)$ to marginalize over the
latent values.
The posterior covariance is approximated by
\begin{eqnarray}\label{approx_redrank_post}
\Sigma\approx \tilde{C}-\tilde{C}(\tilde{C}+W^{-1})^{-1}\tilde{C},
\end{eqnarray}
where $\tilde{C}= \Lambda+\tilde{V}\tilde{S}\tilde{V}^T$ is the
approximate prior covariance with the explicit basis functions
evaluated at the grid points. We can rewrite $(\tilde{C}+W^{-1})^{-1}$
in \eqref{approx_redrank_post} 
as
\begin{eqnarray}
(\tilde{C}+W^{-1})^{-1}=E-E\bm{1}(\bm{1}^TE\bm{1})^{-1}\bm{1}^TE,
\end{eqnarray}
where
\begin{eqnarray}\label{approx_post_comp}
E=D^{1/2}(Z-ZD^{1/2}\tilde{V}\tilde{S}^{1/2}(LL^T)^{-1}\tilde{S}^{1/2}\tilde{V}^T
D^{1/2}Z)D^{1/2}.
\end{eqnarray}
%
In \eqref{approx_post_comp} 
we have denoted $Z=(I_m+D\Lambda)^{-1}$ which is diagonal and
therefore fast to evaluate.
The matrix $L$ in Equation~\eqref{approx_post_comp} is an $(s+5)\times
(s+5)$ lower triangular matrix, and it can be computed using the
Cholesky decomposition
\begin{eqnarray}\label{approx_post_comp_L}
L=\mathrm{chol}(I_{s+5}+\tilde{S}^{1/2}\tilde{V}^TD^{1/2}ZD^{1/2}\tilde{V}\tilde{S}^{1/2}),
\end{eqnarray}
which scales as $\mathcal{O}((s+5)^3)$.
%
%
When drawing samples from the Gaussian approximation
$q(\f|X,\y,\theta)$ given the covariance matrix
\eqref{approx_redrank_post}, the structure of $\tilde{C}$ can be
exploited to avoid forming any $m\times m$ matrix.

With the reduced-rank approximation based on the eigendecomposition,
we avoid choosing (or optimizing) the locations of inducing inputs,
which is required in many sparse approximations, such as, fully
independent conditional (FIC) sparse approaches
\citep[e.g.][]{Snelson+Ghahramani:2006,Quinonero-Candela+Rasmussen:2005}.
With the possibility to choose the locations of inducing inputs, the
reduced-rank approximation would be more expressive although choosing
automatically the locations of inputs can be challenging if the
correlation structure of a GP prior is wanted to be preserved using a
smaller number of inducing inputs. In addition, the optimization of
the locations of inducing inputs is not trivial and can lead to
overfitting
%
%
\citep[see, e.g., the discussion and visualizations of different
correlation structures by][]{Vanhatalo+Pietilainen+Vehtari:2010}.
The problem with the reduced-rank approximation based on Kronecker
products, however, is that the covariance function must be separable
with respect to the inputs, 
which leads to a restricted class of possible covariance functions. In
this paper, we have tested the reduced-rank approximation with the
squared exponential covariance function.

\subsection{Density Regression with the Laplace Approximation}
\label{sec:density_regression}


Logistic Gaussian processes are also suitable for estimating
conditional densities $p(t|\x)$, where $t$ is a target variable
\citep[see, e.g., ][]{Tokdar:2010}. We discretize both the covariate
and target space in a finite region to model the conditional densities
with the logistic GP.
%
%
%
%
In this paper, we focus on modelling the conditional density of $t$
given a univariate predictor $x$, which leads to a 2D grid similarly
as in the case of 2D density estimation.
%
%
%
%
%
%
We denote the number of intervals in the covariate space with $m_x$ and the
number of intervals in target space with $m_t$.
To estimate the conditional densities, we write the log-likelihood
contribution of all the observations as
\begin{eqnarray}\label{likelih_densregr}
\log p(\y|\f) = \sum_{i=1}^{m_x} \left( \y_i^T\f_i - n_i\log
\left(\sum_{j=1}^{m_t} \exp(f_{i,j})\right) \right),
\end{eqnarray}
where $f_{i,j}$ is the $j$'th element of $\f_i$. The $m_t \times 1$
vector $\f_i$ contains all the latent values associated with each
subregion conditioned to the covariate $x_i$, that is, the latent values
associated with the $i$'th slice of the grid.
%
%
Similarly, $\y_i$ is a vector of length $m_t$ and consists of the
number of observations in each subregion of the $i$'th slice of the
grid associated with $x_i$. 
%
%
The vector $\f$ contains all the latent values of the grid and $\y$
all the count observations.
%
%
%
%
To approximate the resulting non-Gaussian posterior distribution, we
use Laplace's method.
The likelihood \eqref{likelih_densregr} results in that $W$ in
equation \eqref{post_cov} is a block-diagonal matrix. Similarly as in
Section \ref{sec:newton}, the $i$'th block of $W$ can be factorized
into $R_iR_i^T$, where
\begin{eqnarray}\label{lik_structure_Rcond}
  R_i=\sqrt{n_i}((\mathrm{diag}(\uu_i))^{\frac{1}{2}}-\uu_i\uu_i^T(\mathrm{diag}(\uu_i))^{-\frac{1}{2}}).
\end{eqnarray}
The total number of observations in the $i$'th slice of the grid is
denoted with $n_i$, and the vector $\uu_i$ is formed as in Equation
\eqref{lik_u}, but by considering only the latent values $\f_i$.
Because the LA approach for the LGP density regression follows closely
to the derivation presented in Section
\ref{sec:laplace-approximation}, the implementation issues are omitted
here.
%
%
The density regression becomes computationally challenging with dense
grids and when applied to a larger number of covariate dimensions $d$, and
therefore, computational speed-ups are required, but we do not
consider these in this paper.



\subsection{Markov Chain Monte Carlo}
\label{sec:mcmc}

MCMC sampling enables approximating the integration over the posterior
distribution without limiting to a Gaussian form approximation.
MCMC can be extremely slow but in the limit of a long run it provides
an exact result by which the accuracy of the LA approach can be
measured.

We approximate the posterior distribution by sampling alternatively
from the conditional posterior of the latent values
 $p(\f|X,\y,\theta)$ by using scaled Metropolis--Hastings sampling \citep{Neal:1998}
 and from the conditional posterior of the hyperparameters
 $p(\theta|\f,X,\y)$ by using the no-U-turn sampler (NUTS)
 \citep{Hoffmann+Gelman:2013}. In the experiments, this
 combination was more efficient than other MCMC methods usually
 used for GPs.
%

A fixed length of MCMC chains was determined by estimating the
convergence and effective sample size for different data sets. 
The convergences of MCMC chains were diagnosed with visual inspection
and the potential scale reduction factor \citep{brooks1998}. The
effective sample size was estimated with Geyer's initial monotone
sequence estimator \citep{geyer1992}.
As the sampling from the conditional posterior of the latent values
was significantly faster, each meta-iteration consisted of hundred
sampling steps for $p(\f|X,\y,\theta)$ and one step for
$p(\theta|\f,X,\y)$. The sampling was initialised by using the
hyperparameter values at the mode of the LA approximated marginal
posterior and the initial latent values were sampled from the Gaussian
approximation given the initial hyperparameter values. The sampling
consisted of 5100 meta-iterations of which 100 iterations were removed
as a burn-in. The effective sample sizes with different data sets and
random number sequences were about 50--100, which gives sufficient
accuracy for the posterior mean of density estimate. For a grid size
of 400, the computation time was about 2200 seconds.
%


%
%
%
%


%

\section{Experiments}
\label{sec:experiments}

In this section, we examine the performance of the Laplace
approximation for the logistic Gaussian process (LA-LGP) with several
simulated and real data sets. We compare LA-LGP to the MCMC
approximation (MCMC-LGP) and to Griffin's (\citeyear{Griffin:2010})
Dirichlet process mixture of Gaussians with common component variance
(CCV-mixture) and different component variance (DCV-mixture). CCV
model assumes that all mixture components have equal variances. DCV
model allows different variances for the components. Computation for
Dirichlet process mixture models is done with Gibbs sampling.
We compare LA-LGP only to advanced Bayesian kernel methods
\citep{Griffin:2010}, since \citet{Tokdar:2007} and \citet{Adams:2009}
have already shown that logistic GP works better than simple kernel
methods (such as the Parzen method). Griffin showed that CCV- and
DCV-mixture models performed equally or better than other default
priors for mixture models, which is why the other priors are excluded
in our comparisons. We do not compare the performance to other MCMC
based LGP approaches by \citet{Tokdar:2007} and \citet{Adams:2009} as
the prior is the same and if using the same grid the difference would
only be in the implemantation speed and convergence speed of the
different MCMC methods.
LA-LGP and MCMC-LGP were implemented using the GPstuff
toolbox\footnote{http://mloss.org/software/view/451/,http://becs.aalto.fi/en/research/bayes/gpstuff/}
\citep{vanhatalo2013}, and CCV/DCV-mixtures were computed using
Griffin's
code\footnote{http://www.kent.ac.uk/ims/personal/jeg28/BDEcode.zip}.

The squared exponential covariance function was employed in all the
experiments with LGP. We also tested Mat{\'e}rn, exponential, rational
quadratic and additive combinations, but the results did not improve
considerably with these different covariance functions.
To ensure that the same prior is suitable for different scales, we
normalized the grid to have a zero mean and unit variance in all the
experiments.
If not otherwise mentioned we used LGP with a grid size 400.
 
The rest of this section is divided into six parts. 
We compare the performances of LA-LGP, MCMC-LGP and CCV/DCV-mixtures
with simulated 1D data in Section \ref{sec:exp_1d}, and with real 1D
data in Section \ref{sec:exp_1d_real}. In Section
\ref{sec:exp_data_grid}, we test how different grid sizes and the
number of data points affect density estimates. We illustrate density
estimation with simulated 2D data in Section \ref{sec:exp_2d}, and
finally, we demonstrate density regression with one simulated data in
Section \ref{sec:exp_dens_regr}.



\subsection{Simulated 1D Data}\label{sec:exp_1d}

\begin{figure*}[!t]
  \centering
  \includegraphics[scale=0.85,clip]{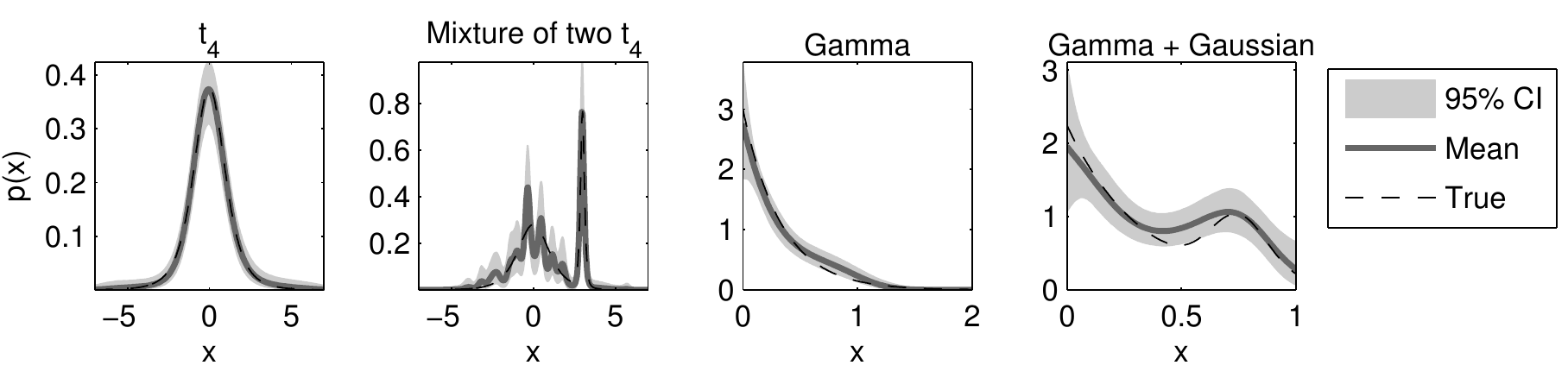}
  \caption{Example results of density estimation and related
    uncertainties for four different simulated data sets with $n=100$.}
\label{fig:sim_examples}
\end{figure*}

Figure \ref{fig:sim_examples} shows the density estimates and the
corresponding 95\% credible intervals for single random realisations
from simulated data sets with $n=100$.
The simulated data sets are:
\begin{itemize}
\item $t_4$: Student-$t_4(0,1)$
\item Mixture of two $t_4$:  $\frac{3}{4} t_4(0,1)+\frac{1}{4}t_4(3,\frac{1}{8^2})$
\item Gamma: $\Gam(1,\frac{1}{3})$
\item Truncated Gamma+Gaussian $x\in(0,1)$: $\frac{3}{4}\Gam(1,\frac{1}{3})+\frac{1}{4}\mathcal{N}(\frac{3}{4},\frac{1}{8^2})$.
\end{itemize}
Student's $t_4$-distribution was chosen as a simple unimodal
distribution with thicker tails than Gaussian. The mixture of two
$t_4$-distributions was chosen as a more challenging case, where there
are two separate modes with different widths. In the second plot of
Figure \ref{fig:sim_examples}, it can be seen that the short
length-scale required to model the narrow peak, makes the density
estimate bumpy also elsewhere. Gamma was chosen as a simple
distribution with a mode on the boundary, and a truncated
Gamma+Gaussian, $x\in(0,1)$, as a more difficult case with one mode on
the boundary and another mode in the middle of the
distribution. Truncated Gamma+Gaussian was also used by \citet{Tokdar:2007} and
\citet{Adams:2009}.

\begin{figure*}[t]
  \centering
  \includegraphics[scale=0.85,clip]{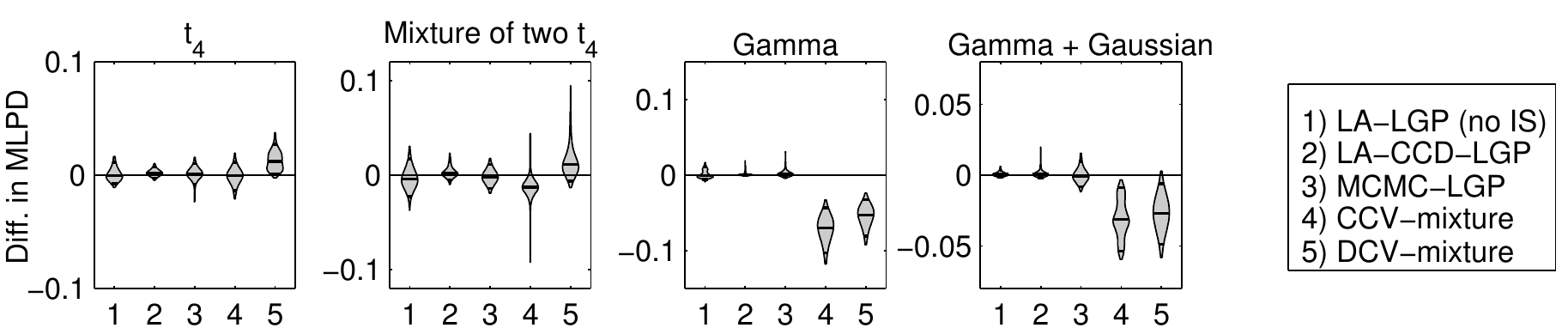}
  \caption{The pairwise comparison of LA-LGP to LA-LGP (no IS), LA-CCD-LGP, MCMC-LGP,
    CCV-mixture and DCV-mixture. The plot shows the distribution of
    differences in the mean log-predictive density (MLPD) along with
    the median and 95\% lines. The values above zero indicate that a
    method is performing better than LA-LGP. The MLPDs were computed
    using the true density as the reference. The violin plots were produced
    using LA-LGP from the results of 100 independent random
    repetitions.  
  }
\label{fig:sim_comp}
\end{figure*}

Figure \ref{fig:sim_comp} shows the pairwise comparison of LA-LGP
to LA-LGP without importance sampling (LA-LGP, no IS), LA-LGP with
the CCD integration (LA-CCD-LGP), MCMC-LGP, CCV-mixture and
DCV-mixture. We made 100 realisations of $n=100$ random samples from
the true density and computed for each method the mean log-predictive
densities (MLPD) over the true distribution. Distributions of the
differences between MLPDs are plotted with violin plots generated
using LA-LGP along with the median and 95\% lines.
There are no statistically significant differences between the methods
for the first three data sets.  For the last data set the Gaussian
mixture models do not work well for the data with the mode on the
boundary. The violin plots in Figures \ref{fig:sim_comp} and
\ref{fig:real_comp} also illustrate another practical application of
density estimation with the LA approach which facilitates interactive
visualisation.

\subsection{Real 1D Data}\label{sec:exp_1d_real}

\begin{figure*}[t]
  \centering
  \includegraphics[scale=0.85,clip]{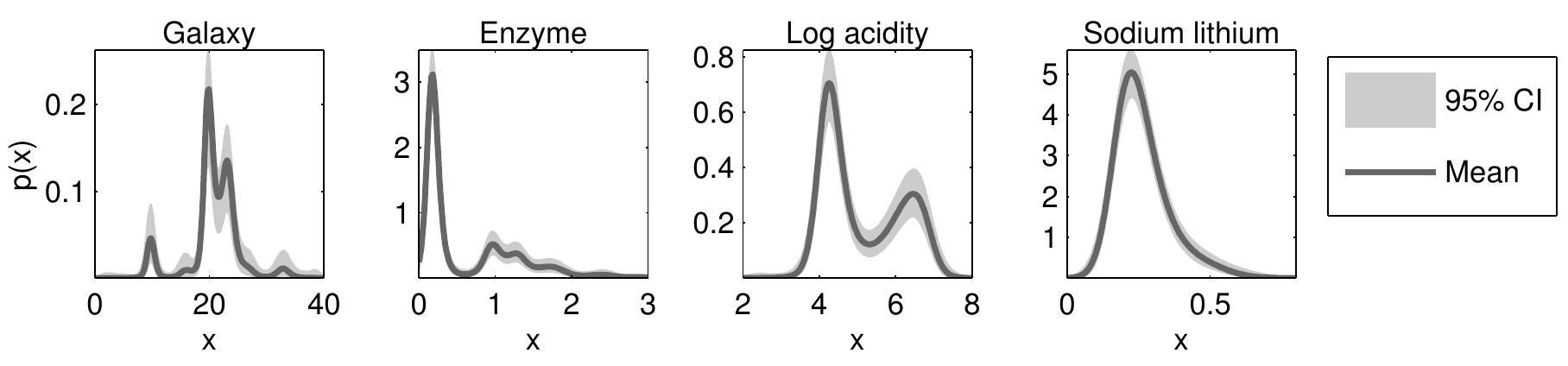}
  \caption{Example results of density estimation and related
    uncertainties for four different real data sets with
    $n=\{82,245,155,190\}$.}
\label{fig:real_examples}
\end{figure*}

Figure \ref{fig:real_examples} shows the density estimates and the
corresponding 95\% credible intervals for the following real data sets:
\begin{itemize}
\item Galaxy $n=82$
\item Enzyme $n=245$
\item Log acidity $n=155$
\item Sodium lithium $n=190$,
\end{itemize}
which were studied by \citet{Griffin:2010}.
The density estimates are visually similar to the results by
\citet[][Figure 5]{Griffin:2010}.

\begin{figure*}[t]
  \centering
  \includegraphics[scale=0.85,clip]{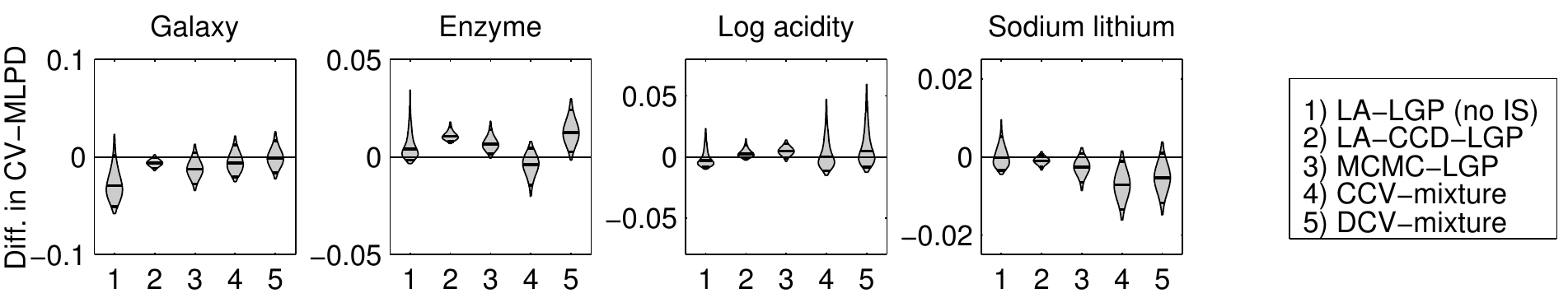}
  \caption{The pairwise comparison of LA-LGP to LA-LGP (no IS), LA-CCD-LGP, MCMC-LGP,
    CCV-mixture and DCV-mixture. The plot shows the distribution of
    differences in the cross-validation mean log-predictive density
    (CV-MLPD) along with the median and 95\% lines. The values above zero
    indicate that a method is performing better than LA-LGP.
    CV-MLPDs were computed using leave-one-out cross-validation.
    Violin plots were produced with LA-LGP given 1000
    Bayesian bootstrap draws from the distribution of CV-MLPD.
  }
\label{fig:real_comp}
\end{figure*}

Figure \ref{fig:real_comp} shows the pairwise comparison of
LA-LGP to LA-LGP (no IS), LA-CCD-LGP, MCMC-LGP, CCV-mixture
and DCV-mixture with the real data sets. For each method, we computed
leave-one-out cross-validation mean log-predictive densities
(CV-MLPD). Samples from the distributions of the differences between
CV-MLPDs were obtained using the Bayesian bootstrap
\citep{Rubin:1981a,Vehtari+Lampinen:2002b} and violin plots were
generated using LA-LGP.

There is no substantial difference in the performances of LA-LGP
and MCMC-LGP across the data sets. 
Importance sampling improves the performance of the Laplace
approximation for the Galaxy data set, but there is no improvement for
the other data sets.
CCD and MCMC improved the performance only for the Enzyme data set.
%
CCV-mixture and DCV-mixture perform similar to LA-LGP for all the
other data sets except for the Sodium lithium data set for which they
have slightly worse performance.

\begin{figure*}[t]
  \centering
  \includegraphics[scale=0.85,clip]{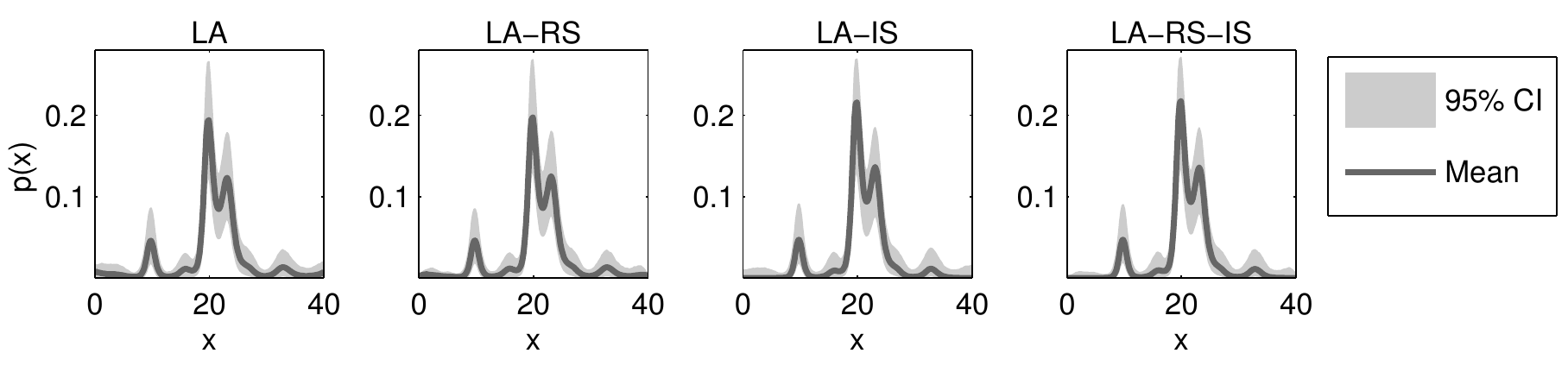}
  \caption{Density estimates for the Galaxy data set with different
    sampling options. The plots show the estimate with the Laplace
    approximation (LA), with additional rejection sampling (LA-RS) or
    importance sampling (LA-IS), or with both rejection and importance
    sampling (LA-RS-IS). See the text for an explanation.}
\label{fig:galaxy_is_vs_rs}
\end{figure*}

Figure~\ref{fig:galaxy_is_vs_rs} shows the effect of the rejection
sampling and the importance sampling in the density estimation of the
Galaxy data set. The rejection sampling helps to make the tails
decreasing. The importance sampling makes the estimate to be lower on
the areas with no observations and respectively higher at the
modes. When looking at MCMC posterior samples, the Galaxy data had
most skewed marginal posterior distributions of the latent values,
explaining the benefit of the importance sampling.


%

\subsection{The Effect of the Number of Data and Grid Points}\label{sec:exp_data_grid}

\begin{figure*}[t]
  \centering
  \includegraphics[scale=0.85,clip]{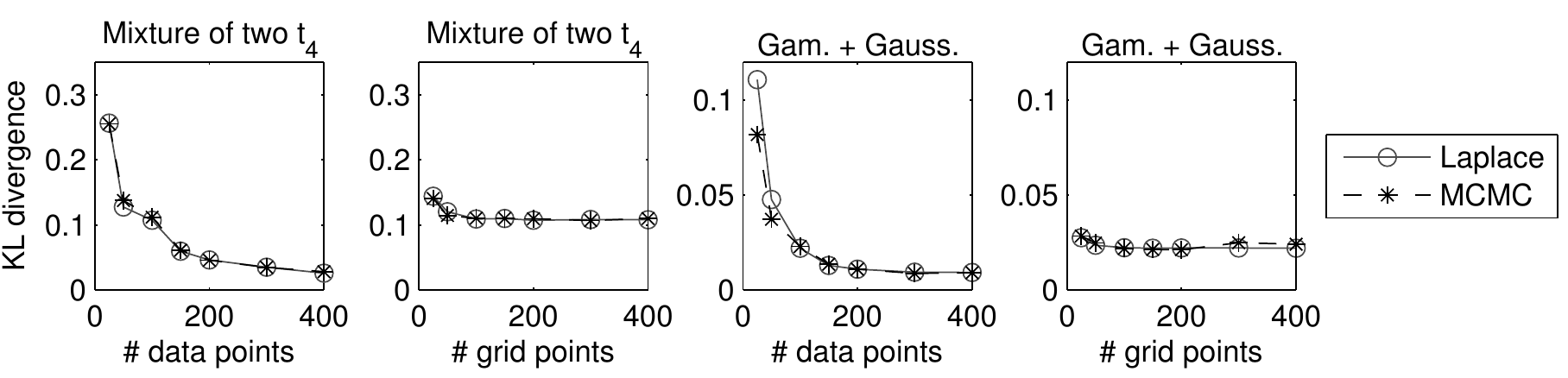}
  \caption{The effect of the number of data and grid points to the
    accuracy of density estimation with two simulated data sets.  The
    number of grid points was 400 when the number of data points was
    varied, and the number of data points was 100 when the number of
    grid points was varied. For both data sets, as the number of data
    points increases, the KL divergence decreases to near zero. About
    50--100 grid points seem to be sufficient for the mixture of
    two $t_4$-distributions, whereas the truncated Gamma+Gaussian distribution is
    less sensitive to the number of grid points. 
    There is no practical difference in the performances of
    LA-LGP and MCMC-LGP.}
\label{fig:n_comp}
\end{figure*}

Figure \ref{fig:n_comp} illustrates the effect of the number of grid
and data points to the accuracy of LA-LGP and MCMC-LGP.  The number of
grid points was 400 when the number of data points was varied, and the
number of data points was 100 when the number of grid points was
varied. For each combination, we made 100 realisations of random
samples from the mixture of two $t_4$-distributions and truncated
Gamma+Gaussian distribution. For both data sets the KL divergence
approaches to zero when the number of data points increase.
It seems that about 50--100 grid points are sufficient for the mixture
of two $t_4$-distributions, whereas the truncated Gamma+Gaussian
distribution is less sensitive to the number of grid
points. 
There is no practical difference in the performances of LA-LGP
and MCMC-LGP.

We measured computation times for density estimation with different
grid sizes using 100 random samples from the Student-$t_4$
distribution. LA-LGP with the grid sizes $(50,100,200,400,900)$
took about $(0.1,0.2,0.3,0.9,4.5)$ seconds using four cores of
Intel(R) Xeon(R) 2.67GHz. If the matrix-vector multiplications in
Newton's algorithm were made with FFT, the times were about
$(0.3,0.3,0.4,0.8,3.4)$ seconds. MCMC-LGP with the same grid sizes
took approximately $(350, 460, 640, 2200, 8100)$ seconds. Using Griffin's
code with the default options, CCV-mixture took about 200 seconds and
DCV-mixture about 800 seconds.
Note that these time comparisons are only approximate, and the
computation times depend considerably on the specific implementation
and on the chosen convergence criteria.

\subsection{Simulated and real 2D data}\label{sec:exp_2d}

\begin{figure*}[t]
  \centering
  \includegraphics[scale=0.85,clip]{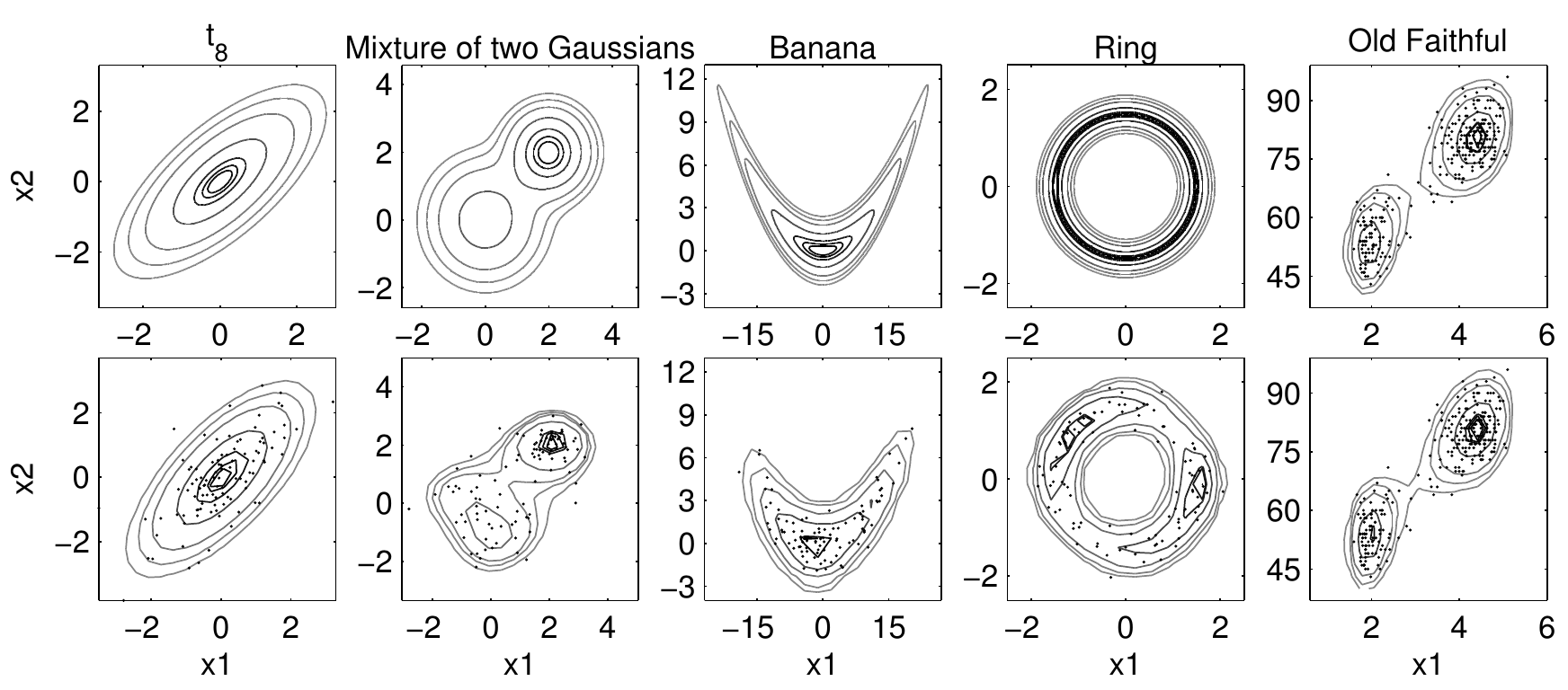}
  \caption{Example results of 2D density estimation for four different
    simulated data sets (Columns 1--4, $n=100$ observations) and for
    one real data set (Column 5, $n=272$ observations). The upper row
    shows the contour plots of the true densities for the simulated
    data sets and the MCMC-LGP result for the Old faithful data. The
    lower row shows the contour plots of the estimated densities
    inferred with LA-LGP. }
\label{fig:2d_examples}
\end{figure*}

The columns 1--4 of Figure \ref{fig:2d_examples} show the following four
simulated 2D distributions:
\begin{itemize}
\item Student-$t_8$: $t_8(0,\begin{bmatrix} 1 & .7 \\ .7 & 1 \end{bmatrix})$ 
\item Mixture of two Gaussians:
  $\frac{1}{2}\mathcal{N}\left(\begin{bmatrix}0\\0\end{bmatrix} ,\begin{bmatrix}1
    &0 \\ 0 & 1 \end{bmatrix}\right)+\frac{1}{2}\mathcal{N}\left(\begin{bmatrix}2\\2\end{bmatrix},\begin{bmatrix}\frac{1}{2}
    &0 \\ 0 & \frac{1}{2} \end{bmatrix}\right)$
\item Banana: $x_1 \sim \mathcal{N}(0,10^2), \quad x_2 \sim \mathcal{N}(\frac{1}{50}x_1^2-\frac{1}{5},1)$
\item Ring:
  $\frac{1}{2\pi}\displaystyle{\int_{-\pi}^{\pi}}\mathcal{N}\left(\frac{3}{2}\begin{bmatrix}\cos\varphi \\\sin\varphi\end{bmatrix},\begin{bmatrix}0.2^2& 0 \\0& 0.2^2\end{bmatrix}\right)d\varphi$.
\end{itemize}
Given a random sample with $n=100$ from each distribution, we compute
the 2D density estimates with LA-LGP. The mean estimates are
illustrated in the lower row of Figure \ref{fig:2d_examples}.
The fifth column of Figure \ref{fig:2d_examples} shows two density estimates for the Old faithful data ($n=272$). The upper plot shows the estimate with MCMC-LGP and the lower plot shows the estimate with LA-LGP.
The density estimation for LA-LGP with a $20\times 20$ grid took about
two seconds and for MCMC-LGP about 29 minutes.

\begin{figure*}[t]
  \centering
  \includegraphics[scale=0.85,clip]{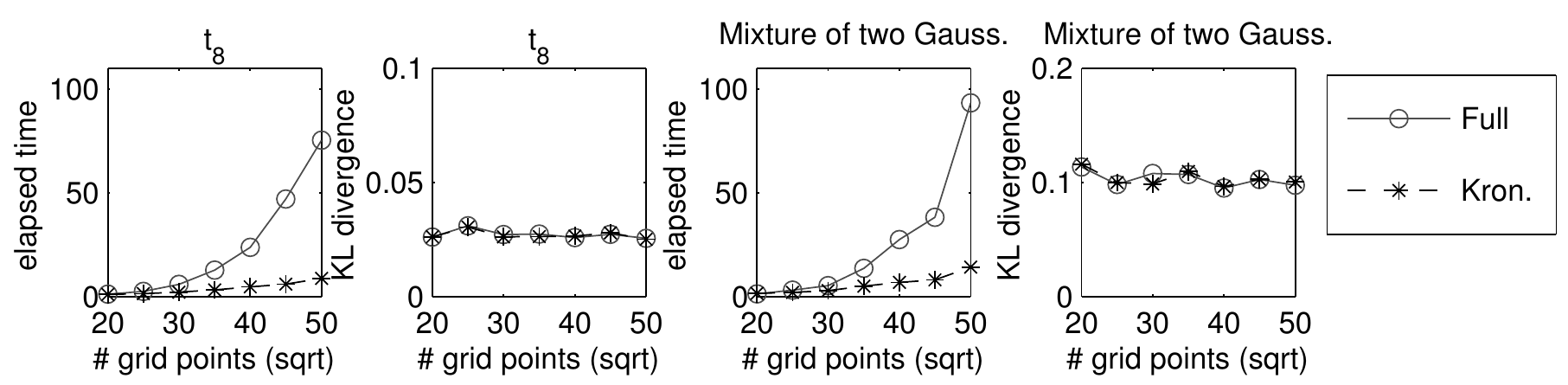}
  \caption{The comparison of LA-LGP with the full prior covariance
    (Full) and with the reduced-rank approximation (Kron.)  of
    Equation~\eqref{approx_prior_cov} for two 2D data sets. The
    performances are similar in the KL sense, but for grid sizes
    larger than $30\times 30$, the computation times are larger with
    the full prior covariance matrix. }
\label{fig:logitgp_2d_kron}
\end{figure*}

We tested the reduced-rank approximation with two 2D data sets. Given
$n=100$ observations from the Student-$t_8$ and mixture of two
Gaussians distributions, we measured the computation times and the KL
divergences from the true density to the estimated density with
different grid sizes.
Figure \ref{fig:logitgp_2d_kron} shows the elapsed times and the KL
divergences as a function of the grid sizes for LA-LGP with the
exact prior covariance (Full) and with the reduced-rank prior
covariance (Kron) of Equation~\eqref{approx_prior_cov}.
%
%
We formed the reduced-rank approximation by excluding all the
eigenvalues smaller than $10^{-6}$, or taking at most 50\% of all the
eigenvalues.
The differences between the exact and the reduced-rank prior are small
in the KL sense, but for the grid sizes larger than $30\times30$,
LA-LGP with the full prior covariance matrix becomes computationally
more expensive than with the reduced-rank approximation.


\subsection{Density Regression with Simulated Data}
\label{sec:exp_dens_regr}

The estimation of a conditional density in a grid with the LA approach
is essentially similar to 1D density modelling with LA-LGP.
Therefore, we consider density regression with only one simulated data
set in this paper.

\begin{figure*}[t]
  \centering
  \includegraphics[scale=0.85,clip]{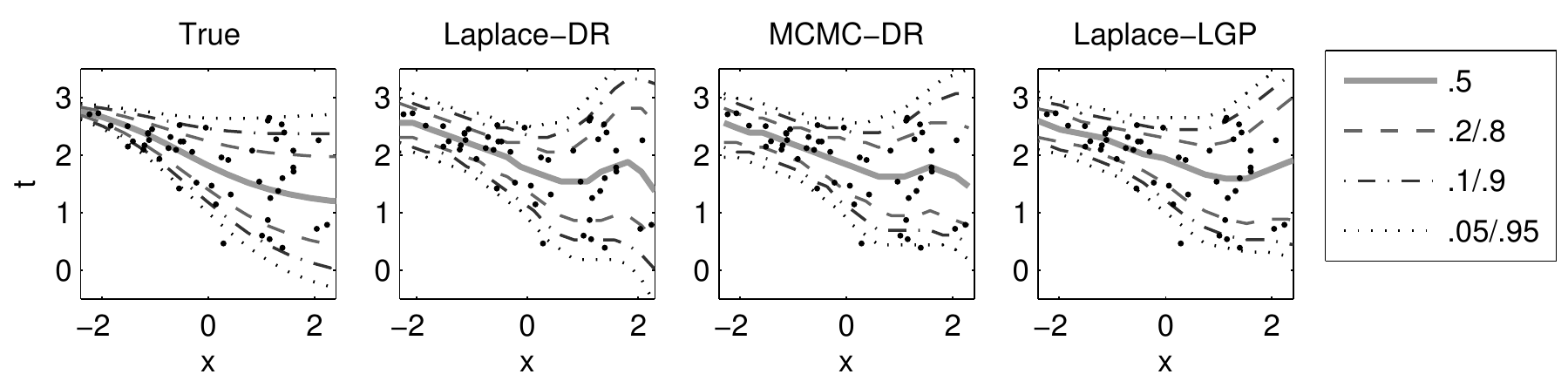}
  \caption{An illustration of density regression with one simulated
    data set. Dots represent a single realisation of $n=50$ samples
    from the simulated density.  The first plot shows the percentiles
    of the true conditional density. The second and third plots show
    the estimates with the Laplace approximation (LA-DR) and with the
    MCMC sampling (MCMC-DR).  The fourth plot shows the conditional
    density estimate computed from a 2D density estimate with LA-LGP.}
\label{fig:density_regression}
\end{figure*}

We demonstrate density regression with a simulated case studied by
\citet{Kundu+Dunson:2011}.  The first plot of Figure
\ref{fig:density_regression} shows the simulated density regression
scheme, where the predictors $z_i$ ($i=1,\ldots,50$) are generated
from the following trimodal density:
$\frac{9}{20}\mathcal{N}(-\frac{6}{5},(\frac{3}{5})^2)+\frac{9}{20}\mathcal{N}(\frac{6}{5},(\frac{3}{5})^2)+\frac{1}{10}\mathcal{N}(0,(\frac{1}{4})^2)$.
The generating model for a univariate target is
$t_i=\lambda\exp\left(-\frac{e^{z_i}}{1+e^{z_i}}\right)+\frac{e^{z_i}}{1+e^{z_i}}\epsilon_i$,
where $\epsilon_i\sim\mathcal{N}(0,\sigma_{\epsilon}^2)$. We fixed
$\lambda=3$ and $\sigma_{\epsilon}=1$.
The second and third plots of Figure \ref{fig:density_regression} show
estimates with the Laplace approximation (LA-DR) and the MCMC
sampling (MCMC-DR) given a single realisation of $n=50$ samples.
Density regression in a $20\times 20$ grid with LA-DR took about
three seconds 
and with MCMC 1600 seconds.
Finally, to show the differences between the estimates obtained with
LA-DR and LA-LGP, we illustrate a conditional density
estimate computed directly from a 2D density estimate obtained with
LA-LGP (the fourth plot of Figure \ref{fig:density_regression}).

\section{Discussion}
\label{sec:discussion}

In this paper, we have proposed to use Laplace's method for fast
logistic Gaussian process density estimation.
%
%
%
%
The empirical results with 1D data sets indicate that the accuracy of
the proposed LA approach with type-II MAP estimation is close to
the state-of-the-art MCMC methods for density estimation. The logistic
Gaussian process with Laplace's method also avoids the sampling and
convergence assessment problems related to the highly multimodal
posterior distributions of the mixture models.

Density estimation with LA-LGP and 400 grid points takes one to two
seconds, which in interactive use is a reasonable waiting time.
For a finer grid or more dimensions, more grid points could be placed,
but this requires additional approximations. In this paper, we have
considered a reduced-rank approximation for LA-LGP that avoids
the infamous cubic scaling of the basic Gaussian process computation.
In addition, we have demonstrated the suitability of the Laplace
approach for estimating conditional densities with one predictor
variable.

Instead of Laplace's method, other deterministic approximations could
be applied to speed up the posterior inference compared to MCMC.
%
%
Variational approximations, including variational bounding and
factorized variational approximations have been considered for GP
models. Although these are close to LA in speed, and can be more
accurate than LA with suitable hyperparameter settings, they can also
have problems in estimating the hyperparameters \citep[see, e.g., the
comparisons between various Gaussian approximations in GP
classification problems by][]{Nickisch+Rasmussen:2008}.
Expectation propagation (EP) has been shown to perform better than the
Laplace or variational approximations for binary classification
\citep{Nickisch+Rasmussen:2008} and for Student-$t$ regression
\citep{Jylanki+Vanhatalo+Vehtari:2011}. However, for the Poisson model
EP was only slightly better than the Laplace approximation
\citep{Vanhatalo+Pietilainen+Vehtari:2010}. Because the Laplace
approximation for the logistic Gaussian process performed almost as
well as the MCMC approximation, we believe that EP could only slightly
improve the performance of LA.
%
The implementation of expectation propagation for non-diagonal $W$ is
non-trivial. A quadrature-free moment matching for EP could also be
considered for density estimation, in a similar way as was done for
multiclass GP classification \citep{Riihimaki+Jylanki+Vehtari:2013},
but in our preliminary testing the moment matching of the
distributions turned out to be quite slow.

%
The Gaussian approximations can be improved by considering corrections
for the marginal posterior distributions \citep{rue2009,cseke2011}.
These corrections can be challenging for the LGP model because the
likelihood function cannot be factorized into terms depending only on
a single latent value.
In the future, it would be interesting to see whether similar
corrections as considered by \citet{rue2009}, could be extended for
LA-LGP where each likelihood term depends on multiple latent values.

%

The code for LA-LGP, MCMC-LGP and violin plots are
available as part of the free GPstuff toolbox for Matlab and Octave
(\url{http://becs.aalto.fi/en/research/bayes/gpstuff/} or \url{http://mloss.org/software/view/451/}).

\subsubsection*{Acknowledgments}

Authors would like to thank Enrique Lelo de Larrea Andrade for his
help in the implementation of the importance sampling, and Pasi
Jyl\"anki, Janne Ojanen, Tomi Peltola, Arno Solin and anonymous
reviewers for helpful comments on the manuscript.
We acknowledge the computational resources provided by Aalto
Science-IT project.
The research has been funded by the Academy of Finland (grant 218248).

\bibliography{logitgp}

\end{document}